\documentclass{PoS}

\usepackage[latin1]{inputenc}                    
\usepackage{graphicx}                            
\usepackage{latexsym}                            
\usepackage{amsfonts}                            
\usepackage{amssymb}                             
\usepackage{amsmath}                             
\usepackage[mathscr]{eucal}                      
\usepackage{dcolumn}                             
\usepackage{latexsym}
\usepackage{epsfig}

\def\det{\mathop{\rm det}}              
\def\ln{\mathop{\rm ln}}                
\def\Dslash{\hbox{D}\kern-0.6em\raise0.15ex\hbox{/}} 

\newcommand{\sgn}{\ensuremath{{\rm sgn}}}

\newcommand{\eps}{\ensuremath{\varepsilon}}

\newcommand{\gf}{\ensuremath{\gamma_{5}}}

\PoS{PoS(LAT2005)146}

\title{Comparison of Chiral Fermion Methods }

\ShortTitle{Comparison of Chiral Fermion Methods }

\author{\speaker{Robert G. Edwards}\thanks{Submitted to PoS on 12th October 2005.}\\
        Jefferson Lab, 12000 Jefferson Ave, Newport News, VA 23606, USA.
        E-mail: \email{edwards@jlab.org}}

\author{B\'alint Jo\'o\thanks{Address since September 16, 2005: Jefferson Lab, 12000 Jefferson Ave, Newport News, VA 23606, USA} \\
        {\em UKQCD Collaboration} \\
	School of Physics, University of Edinburgh, King's Buildings\\
        Edinburgh EH9 3JZ, UK\\	
        E-mail: \email{bjoo@jlab.org}}

\author{Anthony D. Kennedy\\
        School of Physics, University of Edinburgh, King's Buildings\\
        Edinburgh EH9 3JZ, UK\\
        E-mail: \email{adk@ph.ed.ac.uk}}

\author{Kostas Orginos\\
        Jefferson Lab and Department of Physics, College of William and Mary\\
	P.O. Box 8795, Williamsburg, VA 23187-8795, USA \\
        E-mail: \email{kostas@jlab.org}}

\author{Urs Wenger\thanks{Address since 1st October 2005: Institute for Theoretical Physics, ETH Z\"urich, CH-8093 Z\"urich, Switzerland.}\\
        John von Neumann Institut f\"ur Computing NIC\\
        Platanenallee 6, D-15738 Zeuthen, Germany\\
        E-mail: \email{urs.wenger@desy.de}}

\abstract{We present a comparison of various five-dimensional
representations of chiral fermions considering their cost 
and residual chiral symmetry breaking.}

\FullConference{XXIIIrd International Symposium on Lattice Field Theory\\
		 25-30 July 2005\\
		 Trinity College, Dublin, Ireland}

\begin{document}

\section{Introduction}

The realization of chiral symmetry on the lattice represents a major
breakthrough in the field of nonperturbative studies of QCD -- in
particular it allows the numerical simulation of QCD on the lattice
with realistically light fermion flavors. The key to this success is
the overlap Dirac operator which is related to the domain wall
formalism in five dimensions
\cite{Kaplan,Shamir,overlap,neuberger:97a}.
 

This paper provides a comprehensive discussion of various 5D versions of
the overlap Dirac operator where the matrix sign function is approximated
by a rational function.  We show explicitly how the different variants
emerge from the different representations of the rational function and how
they all reduce to the same 4D effective lattice Dirac operator which is the
usual overlap operator or its Hermitian version.
In particular we show that the standard domain
wall formulation is simply just one specific member of a large class of 5D
operators.

An important goal of the numerical work is to evaluate the ``cost''
of various chiral fermion actions. We only consider 5D operators - the ultimate
aim is their use in the force term in Hybrid Monte Carlo simulations.


We will consider 5D formulations to the 4D Overlap Operator or its Hermitian form:
\begin{eqnarray}
D_4(m) &=& \frac{1}{2}\left[ (1+m) + (1-m)\gf \sgn(H) \right] \\
D^{H}_4(m) &=& \frac{2}{1-m}\gf D_4(m) = R\gf + \sgn(H), 
\quad R=\frac{1+m}{1-m}
\end{eqnarray}
where $\sgn(H)$ is approximated by a rational function $\eps(H)$.
There is a four dimensional space of algorithms we can exploit:
\begin{itemize}
\item
choice of kernel $H$,
\item
representation (form) of $\eps(H)$. We consider partial fraction, 
continued fraction and Euclidean Cayley Transform  representations,
\item
approximation of $\sgn(H)\approx \varepsilon_{n,m}(H) =
\frac{P_n(H)}{Q_m(H)}$ for some polynomials $P_n$ and $Q_m$. This
is usually the set of coefficients that fix the particular
approximation in the chosen representation, and
\item
constraints - how are fermions introduced into the path integral.
\end{itemize}



\begin{figure}[t]
\centerline{\epsfig{file=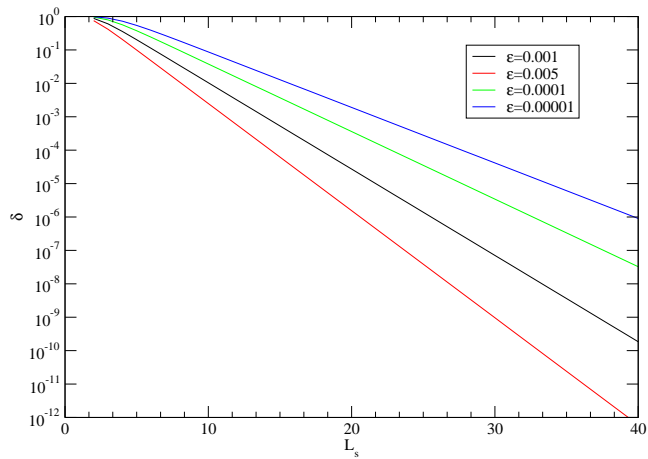,width=3in,angle=-90}}
\caption{
Maximal error of the Zolotarev approximation over the region
$[\epsilon,1]$ as a function of $\epsilon$ and $L_s$.
}
\label{fig:deltavsls}
\end{figure}

Let us consider the various representations of a rational function.  The
Partial Fraction form~\cite{neuberger:98a,edwards:98a, borici:04a} uses
\begin{equation}
\eps_{2n-1,2n}(x) = x\left(p_0 + \sum_{i=1}^{n} \frac{p_i}{x^2 + q_i}\right)
\end{equation}
The Continued Fraction form~\cite{neuberger:99a,borici:01a,Wenger:2004gd} uses
\begin{equation}
\eps_{2n-1,2n}(x) = \beta_0 x +  \frac{1}{\displaystyle \beta_1 x + \frac{ 1 }{\displaystyle \ldots + \frac{1}{\displaystyle \beta_{2n} x}}}
\end{equation}
The Euclidean Cayley Transform form connects the overlap form to the domain
wall form \cite{borici:04a,borici:99a,edwards:00a,chiu:02a,brower:04a}. It uses
\begin{equation}
\eps_{2n-1,2n}(x) = \frac{T^{-1}(x) - 1}{T^{-1}(x)+1}, \quad T(x) = \prod_{i=1}^{2n} \frac{1-\omega_i x}{1 + \omega_i x}
\end{equation}
$T$ is the transfer matrix in the fifth dimension, and the
$\omega_i$ come from the specific approximation. 

There are various approximation functions for $\sgn(H)$.
These approximations give coefficients for all the representations
mentioned above.
One can also supplement the approximation with eigenvectors of the
$H$ operator to decrease errors~\cite{edwards:98a};
projection can be used in all representations~\cite{edwards:00a}.
The original Higham-Tanh approximation \cite{Pandey:1990}
(a.k.a. Neuberger's Polar form~\cite{neuberger:98a}) is
\begin{equation}
\eps_{2n-1,2n}(x) = \tanh \left( -2n \ln \left| \frac{1-x}{1+x} \right| \right) 
  = \tanh(2 n \ \tanh^{-1}(x))= \frac{(1+x)^{2n} -(1-x)^{2n} } {(1+x)^{2n}+(1-x)^{2n}}\quad .
\end{equation}
We note that $\eps_{2n-1,2n}(x) = \eps_{2n-1,2n}(1/x)$.

Over a desired approximation region ($x \in [\epsilon, 1]$), the
approximation due to Zolotarev
\begin{equation}
\eps_{2n, 2n-1}(x) = x R_{n,n}(x^2),
\quad R_{n,n}(x^2) 
 = A \frac{\prod_{l=1}^{n} (x^2 + \bar{c}_{2l})}{\prod_{l=1}^{n}(x^2 + \bar{c}_{2l-1})}
\end{equation}
with
\begin{equation}
\bar{c}_{2l}=-\kappa^2 {\rm sn}^2\left(2iKl/n, \kappa \right) \quad \bar{c}_{2l-1}=-\kappa^2 {\rm sn}^2\left(iK\frac{2l-1}{n}, \kappa \right)
\end{equation}
has much reduced errors for a fixed number of poles compared to the 
``tanh'' approximation. Here, $\kappa=\sqrt{1-\epsilon^2}$ and $K=K(\kappa)$ 
is the complete elliptic integral. The $\omega_j$ are the locations in
$x$ where $\eps_{2n-1, 2n}(x) = 1$.

We plot in Figure~\ref{fig:deltavsls} the maximal
error of the Zolotarev approximation over the interval $[\epsilon,1]$ 
as a function of $L_s$ for various values of $\epsilon$. The maximal error of the tanh approximation
over the same interval is very close to 1 - the maximal error occurs at
$x=\epsilon$.

The kernel $H$ - the auxiliary Dirac operator - is what determines the
``physics'' of the problem. For example, this is where the
discretization errors enter. There are an infinite number of choices
here, but some common variants are as follows. The overlap-like
version is $H = H_w = \gf D_w(-M)$ where $D_w(-M)$ is the Wilson-Dirac
operator with supercritical mass\footnote{Also referred to as the domain wall height} $0<M<2$. In the Shamir domain wall version, the induced operator is
{$H = H_T = H_w ( 2 + a_5 D_w )^{-1}$ }. The M\"obius
version~\cite{brower:04a} $H=(b_5+c_5)H_w [2 + (b_5-c_5)D_w]^{-1}$
interpolates between {$H_w$} and {$H_T$}.

One could use smeared links within the operator to improve the 
spectrum (see e.g.~\cite{Durr:2005an}).
We require  {$H$} to be local, and differentiable (for dynamical fermion
calculations). Ideally, the approximation should cover the spectrum of $H$.
Different choices of {$H$} lead to different spectra and different 
$O(a)$-discretization effects in $H$ and therefore to different $O(a^2)$
effects in the resulting $D_4$ operator.

\section{\bf Introducing the Schur Decomposition}\label{sec:schur}
The framework we shall use for 5D chiral fermion operators is based on
Schur decomposition. Using a 5D form gives the obvious benefit of
inversion of 4D chiral operator within a single (5D) Krylov space.
Furthermore, this framework allows for a formal
reduction of the 5D theory to the 4D theory.
It also allows for a formal derivation of the determinant of the 5D operator
for 5D dynamical fermion calculations. 
Even-odd preconditioning can also be accommodated in this framework.

Consider the 5D block matrix
\begin{equation} 
M_5 = \left(\begin{array}{cc}
A &  B \\
C & D \end{array} \right)
\end{equation}
with $D$ a 4D sub-matrix.
The Schur decomposition of $M_5$ is:
\begin{equation} \label{eq:m5ldu}
M_5 ={\tilde{L}} {\tilde{D}} {\tilde{U}} = \underbrace{\left( 
\begin{array}{cc}
{1} & 0 \\
{CA^{-1}} & {1} \end{array} \right)}_{\displaystyle \tilde{L}} 
\underbrace{\left( \begin{array}{cc} 
{A} & 0 \\
0 & {S} \end{array} \right)}_{\displaystyle \tilde{D}}
\underbrace{\left( \begin{array}{cc}
{1} &{ A^{-1}B} \\
0 & {1} \end{array} \right)}_{\displaystyle  \tilde{U}} \ \mbox{where} \ S = D - CA^{-1}B
\end{equation}
is referred to as the Schur Complement.
The essence of the 5D framework is that $D_4(m)$ (or $D^{H}_4(m)$) is
the 4D Schur Complement of $M_5$ (or of a unitary transformation of
$M_5$). The use of the Schur Complement in 5D formulations was 
advocated in \cite{borici:04a} in connection with multigrid techniques.

The Schur decomposition automatically provides inversion of the 4D system
through inversion of the 5D system. We have
\begin{equation}
M_5 \left( \begin{array}{c}
{\phi} \\
{\psi_4} \end{array} \right) = 
\left( \begin{array}{c}
{ 0} \\
{\chi_4} \end{array} \right) \Rightarrow 
\overbrace{\left( \begin{array}{cc}
 {A} & 0 \\
 0 & {S}
\end{array} \right)}^{\displaystyle {\tilde{D}}}
\overbrace{\left( \begin{array}{c}
{\phi + A^{-1}B \psi_4} \\
 {\psi_4} \end{array} \right)}^{\displaystyle {\tilde{U}} \left( \begin{array}{c} {\phi} \\ {\psi_4} \end{array} \right)} = 
\overbrace{ \left( \begin{array}{c} 
 0 \\
{\chi_4}
\end{array} \right)}^{\displaystyle {\tilde{L}^{-1}} \left( \begin{array}{c}{0}\\ {\chi_4} \end{array} \right) } .
\end{equation}
and it is clear, that solving the 5D system solves the 4D system $S \psi_4 = \chi_4$.

The Schur decomposition  also provides the determinant and the effective 4D operator. By inspection of eq.~(\ref{eq:m5ldu}), we see that
${\det({\tilde{L}})=\det({\tilde{U}})=1} {\Rightarrow \det(M_5) = \det(A)\det(S)}$.
As a consequence, we need an operator that will cancel bulk 5D modes in $M_5$ (i.e., $\det(A)$).
This operator is often called the Pauli-Villars operator which we define
here as follows:
\begin{equation}
M_{PV} = {\tilde{L}} {\tilde{D}_{PV}} {\tilde{U}}, \quad {\tilde{D}_{PV} } = \left( \begin{array}{cc}{A} & 0 \\ 0 & {1} \end{array} \right) \Rightarrow {\det(M_{PV})} ={\det(\tilde{D}_{PV})} = {\det(A)}\quad.
\end{equation}
Formally at least, {$M_5 M^{-1}_{PV}$} is an effective 4D operator:
\begin{equation}
M_5 M^{-1}_{PV} \left( \begin{array}{c}{0} \\ {\psi_4} \end{array} \right) 
={\tilde{L}}\left( \begin{array}{cc}{1} & 0 \\ 0 & {S} \end{array} \right) {\tilde{L}^{-1} } \left( \begin{array}{c}{0} \\ {\psi_4} \end{array} \right) = \left( \begin{array}{c} {0} \\ {S \psi_4 }\end{array} \right)\quad .
\end{equation}

We can now connect the Schur complement formalism to the 5D action and
path integral (the constraints).  Requiring the existence of $A^{-1}$
and the positivity of $S$ only, we have for the partition function:
\begin{eqnarray}
Z &=& \int dU  d{\psi} d{\bar{\psi}}  \ \exp\left\{-{\bar{\psi}} \left({ M_5 M^{-1}_{PV}} \right) {\psi} \right\} = \int dU d{\psi} d{ \bar{\psi}} \ \exp \left\{-{\bar{\psi} \tilde{L}} \left( \begin{array}{cc}{1} & 0 \\ 0 & {S} \end{array}\right){ L^{-1} \psi} \right\} \\
  &=& \int dU d{\chi} d {\bar{\chi}} \ \exp\left\{-{\bar{\chi}}  \left( \begin{array}{cc}{1} & 0 \\ 0 & {S} \end{array}\right){\chi} \right\}, \quad \mbox{defining} \  {\bar{\chi}}={\bar{\psi}\tilde{L}}, {\chi}={\tilde{L}^{-1} \psi} \\
  &\propto& \int dU d{\chi}_4 d{\bar{\chi}}_4 \ \exp \left\{-{\bar{\chi}}_4{S} {\chi}_4 \right\}\quad .
\end{eqnarray}
The path integral needs {$M_{PV}M_5^{-1}$}, but in general {$M_{PV}$} 
involves {$A^{-1}$} (in {$\tilde{L}$}) which can be a dense operator.
However, one can always construct {$\tilde{D}_{PV}$} which needs only {$A$}.
The resulting partition function is
\begin{equation}
Z=\int dU  d\phi^{\dagger} d\phi \ d\eta^{\dagger} d\eta \ \exp\left\{ -\phi^{\dagger} {M_{5}^{-1}} \phi - \eta^{\dagger}{\tilde{D}_{PV}}\eta \right\} = \int dU {\det(A) \det(S)} / {\det(A)} = \int dU \det(S)\ .
\end{equation}
In Hybrid Monte Carlo, one uses {$M^{\dagger}_5 M_5$}, 
{$\tilde{D}_{PV}^{\dagger}\tilde{D}_{PV}$} to ensure integrals converge. Single flavor theories can be simulated with RHMC and taking the square and fourth roots of the squared operators.

\section{Matrix Representations}
We now turn to the specific 5D representations used in this work. For purposes of illustration, we will consider unpreconditioned systems with short 5D extents.
\paragraph{Partial Fraction:}
Let us begin with the partial fraction operator of ref.~\cite{neuberger:00a}.
We show below  partial fraction expansion with only  two poles. The operator is
\begin{equation}
M_5 = \left( \begin{array}{cccc|c}
{H} & {-\sqrt{q_2}} & 0 & 0 & 0 \\
{-\sqrt{q_2}} &{- H} &  0 & 0 & {\sqrt{p_2}} \\
0 & 0 & {H} & {-\sqrt{q_1}} & 0 \\
0 & 0 & {-\sqrt{q_1}} & {-H} & {\sqrt{p_1}} \\
\hline
0 & {\sqrt{p_2}}& 0 & {\sqrt{p_1}} & {R\gf  + p_0  H} \\
\end{array} \right) = \left( \begin{array}{ccc|c} & & & \\ 
& A & & B \\ & & & \\ \hline &  C & & D \end{array} \right)\quad ,
\end{equation}
and the Schur Complement is just the pole approximation to $D^{H}_4(m)$:
\begin{equation}
S = {R\gf + p_0 H} + {\frac{p_1}{H + q_1 H^{-1}}} + {\frac{p_2}{H + q_2 H^{-1}}}
  = R\gf + H \left( p_0 + \sum_{i=1}^{2} \frac{p_i}{H^2 + q_i} \right) \approx D^{H}_4(m)
\end{equation}
and all the results of the section \ref{sec:schur} follow.


\paragraph{Continued Fraction:}
For the continued fraction representation, we  illustrate
the matrix in the $2n+1 = 3$ case:
\begin{equation}
M_5=\left( \begin{array}{ccc}
  {\beta_2 H} & 1 & 0 \\
   1    & {-\beta_1 H} & 1 \\
   0    &   1  & { R\gf + \beta_0 H }
\end{array} \right) 
\end{equation} 
We apply the Schur decomposition recursively. In order to form $S$ we need 
to invert it using its Schur complement $S_1$ and so forth. Finally we arrive at
\begin{equation}
M_5 = \left( \begin{array}{ccc}
         1 & 0 & 0 \\
 { S^{-1}_2} & 1 & 0 \\
  0 &  {S^{-1}_1} & 1 \\ \end{array} \right)
\left(\begin{array}{ccc}
  {S_{2}} & 0 & 0 \\
    0   & {S_{1}} & 0 \\
    0   &  0    & {S} \\
\end{array} \right) 
\left( \begin{array}{ccc}
    1    & {S_{2}^{-1}} & 0  \\
    0    &    1  &  {S_{1}^{-1}} \\
    0    &    0  &   1 
\end{array} \right)
\end{equation}
where
\begin{eqnarray}
{S_2} &=& {\beta_{2}} H, \quad {S_1} = {-\beta_{1}} H - {S^{-1}_2}  \\
{S} &=&{ R \gf + \beta_{0}H} -{S^{-1}_1} = {R\gf + \beta_{0}H} +{( \beta_{1}H + (\beta_2 H)^{-1} )^{-1}} \\
&\approx&
 D^{H}_{4}(m)
\end{eqnarray}
and the outermost Schur complement is $S=D_{4}^{H}$ and all the previous results apply.


\paragraph{Domain Wall:}
The domain wall formalism can be written in a Cayley transform Schur
complement system \cite{borici:04a, borici:99a,edwards:00a,chiu:02a,brower:04a}.
To simplify the formalism, we apply a unitary transformation (defined
in \cite{edwards:00a}) on the domain wall kernel to bring each chiral
half of the physical field onto the same wall. We find that (using a
length of $2n=4$ in the 5th dimension for illustration):
\begin{equation}
Q^{-1}_{-} \gf M_5 \mathcal{P} = \left( 
\begin{array}{ccc|c}
  1        & {-T_{2}^{-1}} & 0 & 0 \\
  0         &    1        & {-T_{3}^{-1}} & 0 \\
  0         &    0        &  1    & {-T_{4}^{-1}} {c_{+}} \\
\hline {-T_{1}^{-1}}  & 0 & 0 & {c_{-}} 
\end{array} \right),
\end{equation}
where
\begin{equation}
{T_{i}} = \frac{1 - {\omega_i}H}{1+{\omega_i} H}, \quad {c_{\pm}} = {\frac{1-m}{2} \pm \frac{1+m}{2} \gf}\quad .
\end{equation}
The Schur complement is:
\begin{eqnarray}
{S} &=& {c_{-}} + {T^{-1}_{1} T^{-1}_2 T^{-1}_3 T^{-1}_4} {c_{+}} , \quad {\rm define} \quad {T^{-1}} = { T^{-1}_{1} T^{-1}_2 T^{-1}_3 T^{-1}_4} \nonumber \\
  &=& {-( T^{-1} + 1 ) \gf} {\left[ \frac{1+m}{2} + \frac{1-m}{2}\gf \frac{T^{-1}-1}{T^{-1}+1} \right]} = {M_{5}(m=1)}{D_{4}(m)}\quad .
\end{eqnarray}

\begin{figure}[t]
\centerline{\epsfig{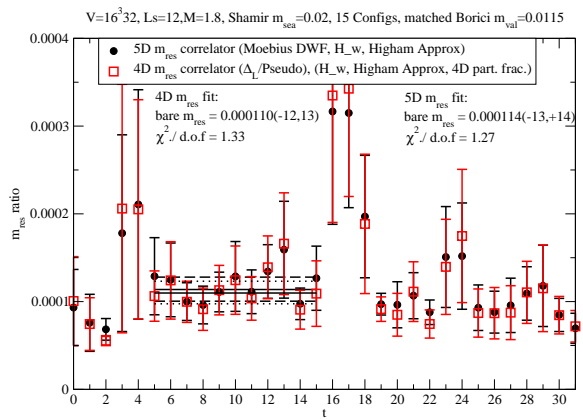}}
\caption{
Comparison of the determination of $m_{res}$ using either a fit
in 4D or 5D to $m_{res}(t)$.
}
\label{fig:fit}
\end{figure}

We find that the domain wall system is a little different from the others.
We see that $\det(A) = 1$ and $S = M_{5}(m=1) D_{4}(m)$.
On the other hand, one can define $M_{PV}=M_{5}(m=1)$ which simplifies its use
and provides a local 5D operator. This result implies
$M_{PV}^{-1}M_{5} = D_{4}(m)$ is an effective 4D operator.
Remaining results then follow from the Schur complement machinery.

\paragraph{Preconditioning and Implementation Tricks:}
Suppose our operator $H$ is a ratio of two operators, $H =
{H_{N}}{H_{D}^{-1}}$ (e.g., the kernel is $H_T$ or $H_{\rm Moebius}$.)
In this case, we naively have an inversion in our 5D operator,
resulting in a nested inversion. We overcome this, by rationalizing
the denominator and solving ${M_5} {H_D H_D^{-1}} \psi = \chi$. We
proceed in two steps, first solving {$({M_5 H_D} )\psi' = \chi$}, then
$\psi={H_D} \psi'$.  After collecting terms in $M_5 H_D$, 
we have just one extra $H$ application to reconstruct $\psi$.

All three formulations allow continuous equivalence transformations on
the approximation coefficients which may be used to make the operator
better conditioned. 

Even-odd preconditioning follows directly from the Schur decomposition
framework developed so far.  The main consideration is the even-odd
block decomposition of the 5D operator\footnote{This is also a Schur Decomposition}. If
\begin{equation}
M_{5} = \left( \begin{array}{cc} M_{ee} & M_{eo} \\ M_{oe} & M_{oo} \end{array}\right) = \left( \begin{array}{cc} 1 & 0 \\  M_{oe} {M^{-1}_{ee}} & 1 \end{array} \right)\left( \begin{array}{cc}M_{ee} & 0 \\ 0 & M_{oo}-M_{oe}{M^{-1}_{ee}} M_{eo} \end{array} \right)\left( \begin{array}{cc} 1 & {M_{ee}^{-1}} M_{eo} \\ 0 & 1 \end{array}\right)
\end{equation}
and if {$M_{ee}$} preserves the structure of {$M_5$}, then the Schur
decomposition of $M_5$ suggests the efficient application of
{$M_{ee}^{-1}$}: {\em one should Schur decompose {$M_{ee}$}.}

For the representations presented here, the efficient use of even-odd
preconditioning depends on $H$. For general $H=H_{\rm Moebius}$, we see that
$M_{ee}^{-1}$ does not depend on the gauge fields and can be
efficiently applied to a vector.


\section{Numerical Results}\label{sec:results}

\begin{figure}[t]
\centerline{\epsfig{file=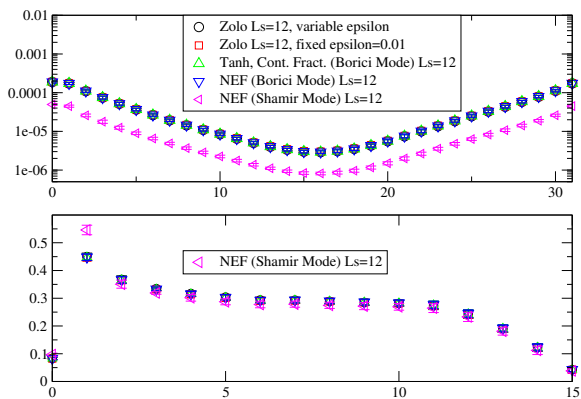,width=3in,angle=0}}
\caption{ Comparison of the pion correlator and effective mass for
various actions, demonstrating how different variants of 5D operators
reduce to the same effective 4D operator. The different kernels $H_w$
and $H_T$ produce different pion correlators, but the pion mass is
tuned to the same value. The difference in mass coming from the
difference between the tanh and Zolotarev coefficients is too small to
show up on these plots.}
\label{fig:pions}
\end{figure}

\subsection{Which 5D Fermion Action?}

We want to evaluate the ``cost'' of various chiral fermion actions.
Here, we only consider 5D inversions for use in the force 
term in HMC and no eigenvector projection is used. We thus have a
residual mass. We choose a winner using a cost metric, which is the number
of Wilson-Dirac applications for fixed residual mass. Using the denominator
rationalization trick mentioned earlier we find that if a general M\"obius DWF operator requires $2n$ applications of $D_{w}$, the corresponding Continued or Partial fraction operators require $2n+1$ such applications. In the case of $H=H_w$, all formulations require only $2n$ applications of $D_w$ (since $\beta_0 = 0$ in the continued fraction and $p_0 = 0$ in the partial fraction case).

\begin{figure}[t]
\centerline{\epsfig{file=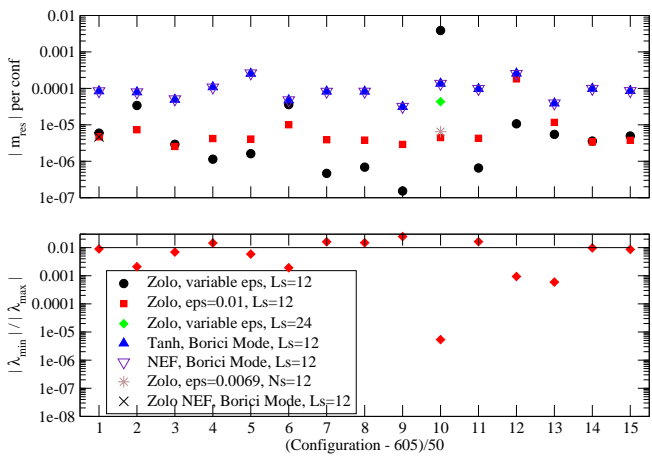,width=3in,angle=-90}}
\caption{
In the upper panel is shown the residual mass per configuration
for various 5D actions using $H=H_w$. Here, ``Zolo'' is the Zolotarev
version of Continued Fraction, ``Tanh'' is the tanh version of Continued
Fraction, ``NEF'' is the tanh domain-wall like variant, and ``Zolo NEF'' is
the Zolotarev domain-wall like variant. In the lower panel is the ratio
of extremal eigenvalues for $H_w$ per configuration.
}
\label{fig:mresconfig}
\end{figure}


\subsection{Residual chiral symmetry breaking}\label{sec:mres}
For our cost comparisons, we note that the
5D definition of the residual mass has been related to the 4D effective
operator~\cite{brower:04a,kikukawa:99a}:
\begin{equation} \label{eq:brower_mres}
m_{\rm res} = {\frac{\sum_{x}\langle \bar{Q}_{x}\gf Q_{x} \ \bar{q}_0 \gf q_{0} \rangle}{(1-m)^{2} \langle \sum_{x} \bar{q}_{x}\gf q_{x} \ \bar{q}_0 \gf q_{0} \rangle}} = \frac{ \sum_x {\rm Tr} \ (D_4^{\dagger}(m))^{-1}_{x,0} \Delta_{x,y} D_4^{-1}(m)_{y,0} }{\sum_{x} {\rm Tr} \ G^{\dagger}_{x,0} G_{x,0} }
\end{equation}
Here, $G = \frac{1}{1-m}(D_{4}^{-1}(m) - 1)$ is the usual subtracted
propagator of the 4D overlap formalism, and $\Delta$ is the breaking term
in the Ginsparg-Wilson relation from the approximation where $\eps \ne
\sgn$:
\begin{equation}
{2 \gf \Delta} {= \gf D_{4}(0) + D_{4}(0) \gf - 2 D_{4}(0)\gf D_{4}(0)} \Rightarrow {\Delta = \frac{1}{4}\left( 1 - \eps^2(H) \right)} \quad .
\end{equation}
where we have used $D_{4}(0)=\frac{1}{2}(1 + \gf \eps(H))$.
We see that $m_{\rm res} \ne 0$ is purely a deficiency of the 
approximation $\eps(H) \approx \sgn(H)$. Further, the numerator 
in eq. (\ref{eq:brower_mres}) can
easily be computed in 4D using the pole approximation of {$\eps(H)$}.
Thus, we can compare chiral breaking for {\em all} formulations on an
equal footing.

Summation is carried out over both spatial and temporal coordinates in
(\ref{eq:brower_mres}). If one does not sum over time, the resulting
time-sliced correlation function is {\em not} identically equal to the
usual Domain Wall $m_{\rm res}(t)$. However, since $\Delta_{x,y}$ is a
local operator, at low energies the two are very similar as can be
seen by observing fig. \ref{fig:fit}. Further, at the cost of
introducing a small contact term on the source time-slice and a
term that vanishes in the ensemble average one may replace the
$D^{-1}$ terms in eq. (\ref{eq:brower_mres}) by the subtracted
propagators $G$. In our computations, this simplification was used,
however, apart from figure \ref{fig:fit} we always carried out the
temporal sum.

\subsection{Setup}

For our numerical tests, we used 15 $N_f=2$ domain-wall fermion
configurations (provided by RBC~\cite{rbc:04a}) with $L_s=12$ and $a
m_q=0.02$ corresponding to $m_\pi=500$MeV.  We compare the cost of
inversions for even-odd preconditioned M\"obius variants of 5D Partial
Fraction, Continued Fraction, and Domain Wall operators. We use only a
CGNE inverter.  Unfortunately, BiCG-stab and MR are not convergent.
In all tests, we used a target residual of $10^{-7}$ and $M=1.8$ in
$D_w$. We tuned the pion mass of our 5D operators using $H=H_w$ to
match the domain-wall pion mass which resulted in an $a m_q=0.0115$.

We computed the cost (number of Wilson-Dirac applications) for
multiple $L_s$ with the tanh and Zolotarev approximations. For the
latter, we found the largest eigenvalue of $H_T$ and $H_w$ to be about
$1.6$ and $5.8$, respectively. We set the upper bound of the
approximation to this $\lambda_{max}$, and set the lower bound to a
{\em fixed} $\epsilon \times \lambda_{max}$ where $\epsilon=0.01$ in
practice.


\begin{figure}[t]
\centerline{\epsfig{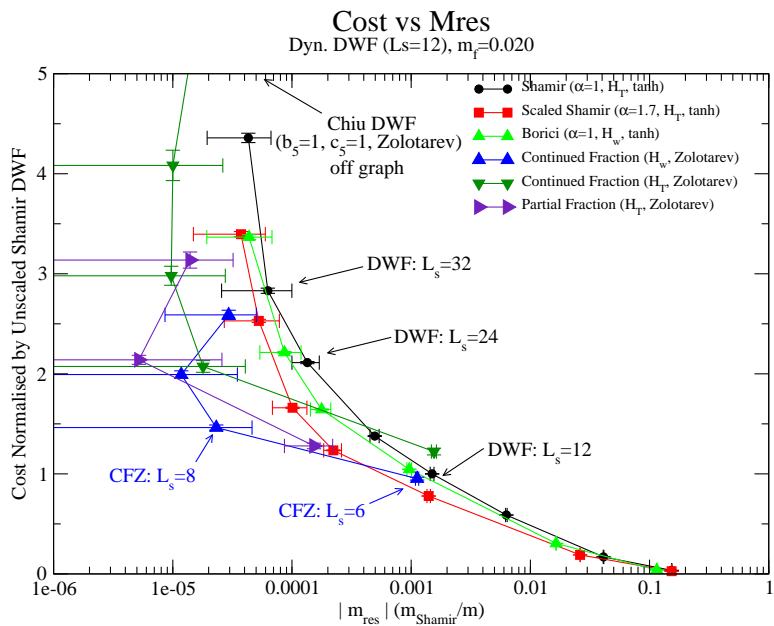}}
\caption{
Cost in Wilson-Dirac applications vs. $m_{\rm res}$ for the most promising
5D actions.
}
\label{fig:cost}
\end{figure}

\subsection{Equivalence of representations/variants}

In Figure~\ref{fig:pions}, we show the pion correlator for different
representations and approximations at the tuned pion mass. We see
that the deviations from each other are small and under control.

\begin{figure}[t]
\centerline{\epsfig{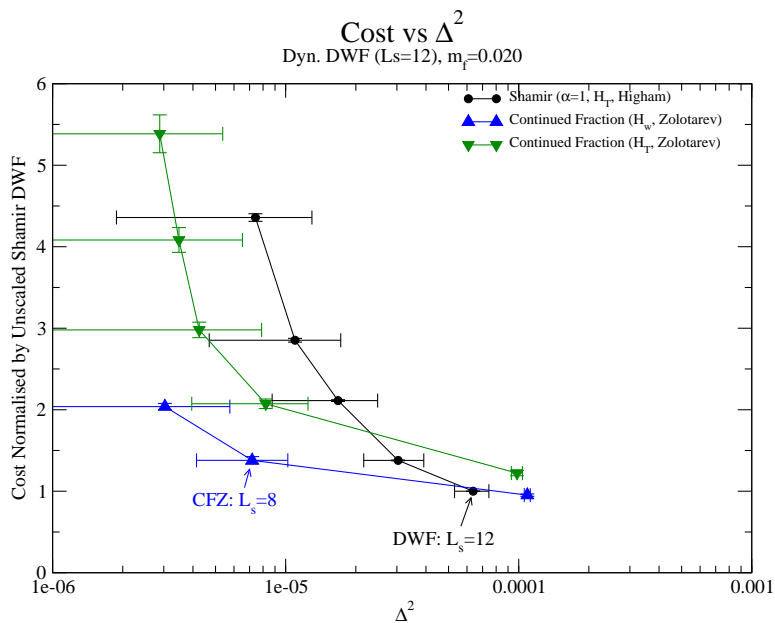}}
\caption{
Cost in Wilson-Dirac applications vs. $\Delta^2$
for some of the 5D actions in Figure~\protect\ref{fig:cost}.
}
\label{fig:cost2}
\end{figure}

\subsection{$m_\text{res}$ per configuration}

In Figure~\ref{fig:mresconfig} we show a plot of $m_\text{res}$ per
configuration for various actions using $H=H_w$. We see that the
``tanh'' versions of domain-wall and continued fraction agree as
expected.  Also, the Zolotarev versions of domain-wall and continued
fraction agree for the first configuration. We did not measure the
Zolotarev version of the domain wall for the remaining configurations
as the cost appeared too expensive, however, we expect similar
agreement we see for the first configuration.  We also see that in the
``tanh'' approximation cases the $m_{res}$ is notably higher than in
the Zolotarev cases. This is consistent with the much larger
approximation errors $\eps(x)$ in polar form compared to Zolotarev.

The notable exception to a small $m_{res}$ is configuration 10 where
the lowest eigenvalue of $H_w$ is quite small. We find that fixing the
approximation range is preferable to allowing it to vary. What we are
seeing is an interplay of the density of eigenvalues of $H_w$ and the
approximation accuracy. The eigenvectors of $H_w$ are also
eigenvectors of $\Delta$ which can be written as $\int d\lambda
\rho(\lambda) \Delta_L(\lambda) = (1/4)\int d \lambda
\rho(\lambda)(1-\eps(\lambda)^2)$. If the density of small modes is
low, it is not worth increasing the approximation range to accommodate
the low modes. Increasing the approximation range increases the maximum error
which thus magnifies $\Delta$ for all modes - not just the smallest
modes. In our tests we therefore fixed the maximal range of the
approximation. We note that is {\em no worse} than a ``tanh''
approximation where the approximation is fixed for a given order of the rational approximation.



\subsection{Cost versus $m_\text{res}$}

\newcommand{\be}{\begin{displaymath}}
\newcommand{\ee}{\end{displaymath}}

In our tests we compare the cost of
inverting the even-odd preconditioned 5D operators
versus  $m_{\rm res}$. We note that this preconditioning gives between a factor of 1.5-3 speedup over the unpreconditioned systems. The cost of
applying $M_{ee}^{-1}$ is negligble in flops; however, some care is needed 
in writing optimized codes due to memory loads in $M_{ee}$. The goal
is small $m_{\rm res}$ for small cost.

In Figure~\ref{fig:cost} we plot the cost versus $m_{\rm res}$ for the
most promising actions. Many of the Zolotarev variants were very
expensive and not shown. We remark, however, that we did not explore
possible further tuning of the condition of our operators via
equivalence transformations on our coefficients, which may reduce the
cost of these expensive variants. Having said this, of all the actions
we tested, the standard domain-wall ($H_T$) is the least
effective. The Zolotarev Continued Fraction ($H_w$ and $H_T$) are the
top candidates. The Zolotarev Partial Fraction version is promising.
We note that rescaling the coefficients $b_5+c_5$ in $H_{\rm Moebius}$
with $b_5-c_5=1$ is equivalent to a rescaling $\alpha=b_5+c_5$ in
$\eps(\alpha H_T)$. For the tanh approximation, we can slide the $H_T$
down inside $\eps(H_T)$ to lower approximation errors. We see this
rescaling results in a corresponding decrease in $m_{\rm res}$ at no
cost increase.

In Figure~\ref{fig:cost2} we plot the cost versus the 
second moment $\Delta^2$. We see no large shift in cost corresponding
to wild oscillations that cancel in $m_{\rm res}$ but not in $\Delta^2$.

\section{Summary and conclusions}

We have presented a unified framework for 5D chiral operators.
The crux of the framework is the Schur decomposition and Schur complement
techniques.
The framework is agnostic about representation, approximation or the kernel.
The framework accommodates (and can help with) even-odd preconditioning.
In practice, the computational cost of a 5D formulation depends upon 
the spectral properties of $M_5$ and upon $m_{\rm res}$.
While $m_{\rm res}$ is an artifact of the approximation $\eps(H)$, 
it depends not only on how good the approximation is, but
also how much of the spectrum of $H$ is {\em not} well approximated 
(e.g., the near-zero modes of $H$).
The condition of $M_5$ depends upon its matrix structure and coefficients,
e.g., the approximation and representation and potentially on further 
tuning of the condition through equivalence transformations and 
permutation symmetry.


We have presented a detailed comparison of the ``cost'' of various
chiral fermion actions, where cost is the number of Wilson-Dirac
applications for fixed $m_{res}$ in a conjugate gradient solver.
We have found the standard domain-wall action is the least effective.
The Zolotarev versions of Continued Fraction (with $H=H_w$ and $H_T$)
appear to be among the best. Future work will focus on dynamical fermion 
implementations of this operator.

\acknowledgments
We gratefully acknowledge support for this work from DOE contract
DE-FC02-94ER40818 and DE-AC05-84ER40150 under which the Southeastern
Universities Research Association operates the Thomas Jefferson
National Accelerator Facility (RGE), from PPARC under grant number
PPA/G/O/2002/00465 (BJ) and from the European Commission's Research
Infrastructures activity of the Structuring the European Research Area
programme, contract number RII3-CT-2003-506079 (HPC-Europa) (UW).
We gratefully acknowledge valuable discussions with Richard Brower.


\end{document}